\documentclass[letter]{aa}
\usepackage{graphicx}
\usepackage{txfonts}
\begin{document}

   \title{Nitrogen fractionation towards a pre-stellar core traces isotope-selective photodissociation}

\author{S. Spezzano\inst{1}  \and P. Caselli\inst{1} \and O. Sipil\"a\inst{1} \and L. Bizzocchi\inst{2}}  

\institute{Max-Planck-Institut f\"ur Extraterrestrische Physik, Giessenbachstrasse 1, 85748 Garching, Germany \and University of Bologna, Dipartimento di Chimica “Giacomo Ciamician”, Bologna, Italy}

\abstract 
{Isotopologue abundance ratios are important to understand the
evolution of astrophysical objects and ultimately the origins of
a planetary system like our own. Being nitrogen a fundamental ingredient of pre-biotic material, understanding its chemistry and inheritance is of fundamental importance to understand the formation of the building blocks of life.}{We aim at studying the $^{14}$N/$^{15}$N ratio in HCN, HNC and CN across the prototypical pre-stellar core L1544. This study allows us to test the proposed fractionation mechanisms for nitrogen.} {We present here single-dish observations of the ground state rotational transitions of the $^{13}$C and $^{15}$N isotopologues of HCN, HNC and CN with the IRAM 30m telescope. We analyse their column densities and compute the $^{14}$N/$^{15}$N ratio map across the core for HCN. The $^{15}$N-fractionation of CN and HNC is computed towards different offsets across L1544.}{The $^{15}$N-fractionation map of HCN towards a pre-stellar core is presented here for the first time. Our map shows a very clear decrease of the $^{14}$N/$^{15}$N ratio towards the southern edge of L1544, where carbon chain molecules present a peak, strongly suggesting that isotope-selective photodissociation has a strong effect on the fractionation of nitrogen across pre-stellar cores. The $^{14}$N/$^{15}$N ratio in CN measured towards four positions across the core also shows a decrease towards the South-East of the core, while HNC shows opposite behaviour. We also measured the $^{12}$CN/$^{13}$CN ratio towards four positions across the core.}{The uneven illumination of the pre-stellar core L1544 provides clear evidence that $^{15}$N-fractionation of HCN and CN is enhanced toward the region more exposed to the interstellar radiation field. Isotope-selective photodissociation of N$_2$ is then a crucial process to understand $^{15}$N fractionation, as already found in protoplanetary disks. Therefore, the $^{15}$N-fractionation in pre-stellar material is expected to change depending on the environment within which pre-stellar cores are embedded. 
The $^{12}$CN/$^{13}$CN ratio also varies across the core, but its variation does not affect our conclusions on the effect of the environment on the fractionation of nitrogen. Nevertheless, the interplay between the carbon and nitrogen fractionation across the core warrants follow-up studies.}
{}

\titlerunning{}
\maketitle

%

\section{Introduction}
Isotopologue abundance ratios are pivotal for tracing the origin and evolution of the molecular material in the process of star and planetary system formation. Stable isotope ratios can in fact be measured in star-forming regions as well as in the Solar System. 
Nitrogen has two stable isotopes, $^{14}$N and $^{15}$N. The $^{14}$N/$^{15}$N ratio varies greatly across the Solar System, depending on material and molecular tracer. This ratio is $\sim$440 in the solar wind and Jupiter (reflecting the composition of the protosolar nebula), $\sim$270 on Earth (for molecular nitrogen), $\sim$150 in comets (for CN, HCN and NH$_2$), and $\sim$200 in protoplanetary disks (for HCN) \citep{furi15, guzman15}.
The origins of the different $^{14}$N/$^{15}$N ratios among Solar System bodies are not yet fully understood. However, given that nitrogen is generally enriched in $^{15}$N in more pristine material like comets, the different $^{14}$N/$^{15}$N ratios that we observe are potentially inherited from the early phase in the formation of the Solar System.
Furthermore, recent observations of different $^{14}$N/$^{15}$N ratios in HCN and CN towards a protoplanetary disk point to the existence of multiple isotopic reservoirs of nitrogen for forming planets \citep{hily-blant19}.
In the dense interstellar medium, the $^{14}$N/$^{15}$N ratio is inferred from N-bearing species and it is 300-500 for CN, HCN and HNC \citep{hily-blant13a, hily-blant13b}. The $^{14}$N/$^{15}$N ratio in N$_2$H$^+$ instead varies from 180 to 1000 \citep{fontani15, bizzocchi13, redaelli18, colzi19}. State-of-the-art reaction networks fail to reproduce the $^{14}$N/$^{15}$N variation in N$_2$H$^+$ \citep{roueff15, wirstrom18}, although a faster recombination with electrons of the $^{15}$N isotopologues in L1544 with respect to the normal species might explain the depletion of N$_2$H$^+$ in $^{15}$N \citep{ loison19,hily-blant20, redaelli20}. The use of $^{13}$C isotopologues to derive the column densities of CN, HCN and HNC is suggested to be done with caution by \cite{roueff15} because of the possible depletion of $^{13}$C with respect to the interstellar medium (ISM) $^{12}$C/$^{13}$C ratio, and of the interdependence of the $^{13}$C and $^{15}$N chemistry \citep{colzi20}. 

Aside from local nuclear synthesis that cause an increasing excess of $^{15}$N towards the Galactic centre \citep{adande12}, two main mechanisms are responsible for the nitrogen fractionation in the ISM: isotope exchange at low-temperatures and isotope-selective photodissociation. Isotope exchange is the main chemical path to enrich molecules in deuterium towards the center of pre-stellar cores (e.g. \citealt{caselli03}). In the case of HCN, as an example for nitrogen bearing species, the isotope exchange reaction is

\begin{equation}
   \rm HC^{14}NH^+ + ^{15}N \rightleftarrows HC^{15}NH^+ + ^{14}N + 35 K,
\end{equation}

\noindent followed by the dissociative recombination of the cation with electrons. The isotope exchange reactions enrich molecules in the heavier isotope at low temperature because of the lower zero-point energy of the heavier molecular species and the exothermicity of the reaction \citep{terzieva00}. The isotope exchange reactions involving nitrogen were later found to have barriers \citep{roueff15}, and hence are not favourable to reproduce its fractionation \citep{wirstrom18}.
Isotope-selective photodissociation favours the photodissociation of the $^{15}$N-bearing isotopologues of N$_2$. Being $^{14}$N$_2$ more abundant, it can self-shield better than $^{14}$N$^{15}$N and is consequently less affected by photodissociation \citep{heays14, visser18}.
With $^{14}$N$^{15}$N more photodissociated than $^{14}$N$_2$, the $^{15}$N available to form molecules increases. The effect of selective isotope-photodissociation on nitrogen fractionation has already been clearly observed in protoplanetary disks \citep{hily-blant19}.
The effect of the local physical conditions on the fractionation of nitrogen in star-forming regions has been recently explored by \cite{colzi18} who found  large scatter in the $^{14}$N/$^{15}$N ratio in HCN towards high-mass star forming cores. However, the spatial variation of the $^{14}$N/$^{15}$N ratio towards pre-stellar cores has not been mapped yet.\\
Isotope-selective photodissociation is expected to be effective only at very low extinction (A$_V$ = 1-3 mag, \citealt{heays14}), and it is not considered in chemical models for the fractionation of nitrogen in dense cores. Its effects are only considered as inherited from the parent cloud \citep{furuya18a, furuya18b}, and they still cannot reproduce the "anti-fractionation" of nitrogen measured in N$_2$H$^+$ \citep{redaelli20}.

In this paper we present the first $^{14}$N/$^{15}$N ratio map towards the prototypical pre-stellar core L1544, showing clear evidence that isotope-selective photodissociation is affecting the $^{15}$N fractionation in HCN. The same result is found for CN towards four offsets across the core, and qualitatively for HNC. In Section 2 we describe the observations and in Section 3 the analysis of the data. Our results are reported in Section 4, and finally we discuss our conclusions and outlook in Section 5.

\begin{figure*}

\begin{center}
\includegraphics[width=15cm]{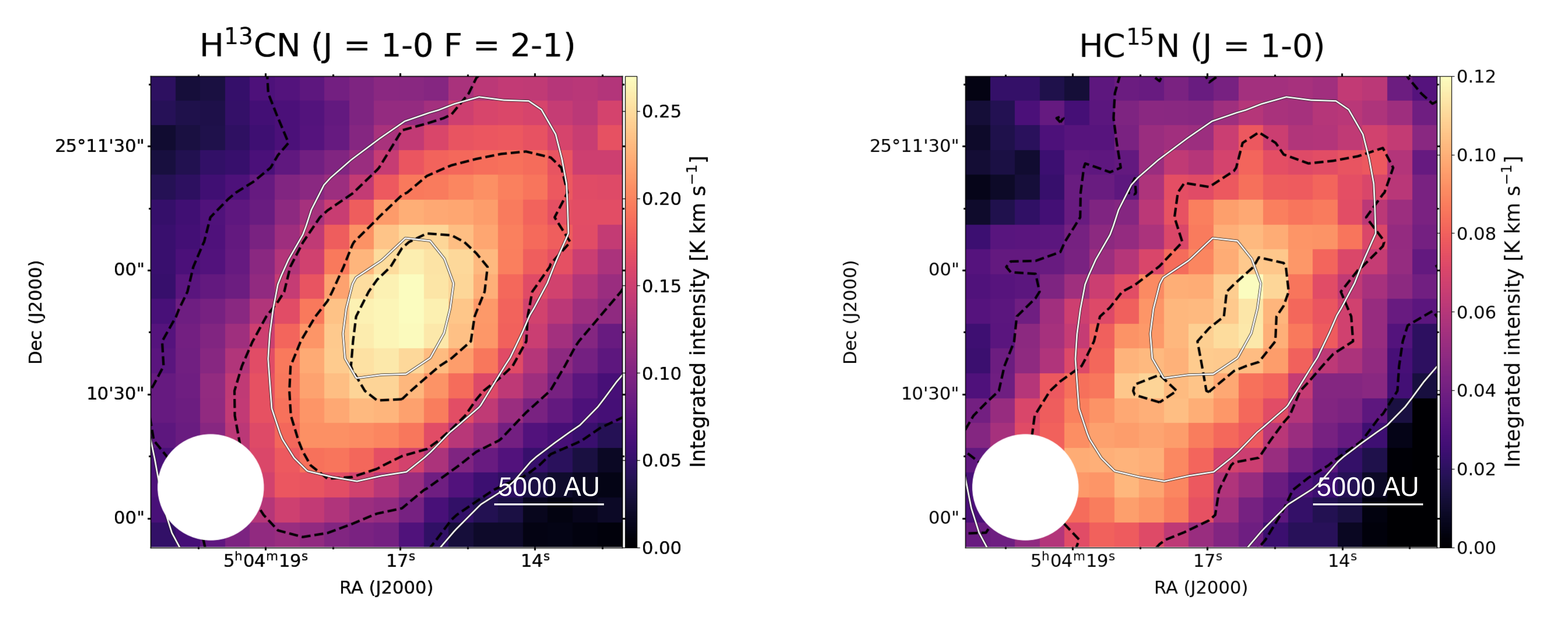}
\end{center}
\caption{Integrated intensity maps of the 1-0 transitions of H$^{13}$CN and HC$^{15}$N towards the inner 2$^\prime$ $\times$2$^\prime$ of L1544. The 30$^{\prime\prime}$ beam of the 30m telescope is shown at the bottom left of each map. The solid white contours are the 30\%, 60\% and 90\% of the peak intensity of the N(H$_2$) map of L1544 computed from $Herschel$/SPIRE data \citep{spezzano16}. The dashed black contours indicate the 10$\sigma$ integrated emission with steps of 10$\sigma$ for H$^{13}$CN, and 5$\sigma$ with steps of 5$\sigma$ for HC$^{15}$N ($rms_{H^{13}CN}$= 6 mK km s$^{-1}$, $rms_{HC^{15}N}$=7 mK km s$^{-1}$).}
\label{fig:ii_maps}
\end{figure*}

\begin{table*}{}
\caption{Spectroscopic parameters of the observed lines}
\label{table:parameters}
\begin{tabular}{ccccccccc}
\hline\hline \\[-2ex]
Molecule & Transition & Rest frequency & $E_\text{up}$ & T$_{MB}$ & v$_{LSR}$  &   FWHM   & W & N$_{tot}$\tablefootmark{a}    \\
&       &(MHz)   & (K)  & K & km/s & km/s & K km/s              & 10$^{12}$cm$^{-2}$                       \\[0.5ex]
\hline \\[-2ex]
H$^{13}$CN   & $J$= 1-0 $F$ = 2-1 &  86340.1666(1)     & 4.1 & 0.53(6) &7.14(2)&0.51(4)&0.28(2)& 3.19(2)   \\
HC$^{15}$N      & $N$= 1-0  &   86054.9664(1)      &  4.1 & 0.235(6) & 7.18(1)& 0.49(1) & 0.120(2)   & 0.56(3)    \\
HN$^{13}$C  &  $N$= 1-0 $F'_1$-$F_1$ = 1-0&    87090.675(4)\tablefootmark{b}        &  4.2   &0.41(1) & 7.143(3) & 0.25(2) &0.10(1) &  2.5(2)     \\
H$^{15}$NC   &  $N$= 1-0 &   88865.676(5)\tablefootmark{c}      &  4.3  & 0.67(2) & 7.257(4) & 0.425(8) & 0.302(7)  &          1.0(2) \\
CN   & $N$= 1-0 $J$ = 1/2-1/2 $F$= 1-1  &   113123.370(5)        &  5.4  &0.87(2) & 7.109(4) & 0.465(8) & 0.427(6)    &517(7)     \\
$^{13}$CN   & $N$= 1-0 $J$ = 3/2-1/2 $F_1$= 2-1 $F$= 3-2  &    108780.20(5)        &  5.2 & 0.13(8) & 7.18(1) & 0.39(3) & 0.057(3)  &2.7(1)      \\

C$^{15}$N   & $N$ = 1-0 $J$ = 3/2-1/2 $F$= 2-1&   110024.6(1)        & 5.3  & 0.056(1)  & 7.09(2) & 0.38(6) & 0.023(4)  &0.48(6)      \\

\hline
\end{tabular}
\tablefoot{Numbers in parentheses denote $1\sigma$ uncertainties in unit of the last quoted digit. The frequencies in the Table are reported in the CDMS and JPL catalogues \citep{muller05, pickett98} and are derived from the laboratory work in \citealt{fuchs04, creswell76, vandertak09, bogey84,klisch95} and the interstellar detection reported in \citealt{saleck94}.
\tablefoottext{a}{Column densities computed towards the dust peak assuming T$_{ex}$ =3.5 K for HCN isotopologues, T$_{ex}$ =4.5 K for HNC isotopologues, and T$_{ex}$ =4.2 K for CN isotopologues \citep{padovani11, hily-blant13a, hily-blant13b}.}
\tablefoottext{b}{One of the four "effective" hyperfine components following the description of \cite{vandertak09}}
\tablefoottext{c}{The rest frequency reported in the JPL catalogue for the 1-0 transition of H$^{15}$NC is 88865.71(4). For this work we used the rest frequency derived from our observations, which is consistent with the JPL value within its error bar. }}
\end{table*}

\section{Observations}
 The emission maps of the 1-0 transition of the $^{13}$C and $^{15}$N isotopologues of CN, HCN an HNC towards L1544 were obtained using the IRAM 30m telescope (Pico Veleta, Spain) in 2 different observing runs in 2013 and 2015. 
 We performed a 2.5$^\prime$ $\times$2.5$^\prime$ on-the-fly (OTF) map centred on the source dust emission peak ($\alpha _{2000}$ = 05$^h$04$^m$17$^s$.21,  $\delta _{2000}$ = +25$^\circ$10$'$42$''$.8). We used position switching with the reference position set at (-180$^{\prime \prime}$,180$^{\prime\prime}$) offset with respect to the map centre. The observed transitions are summarised in Table \ref{table:parameters}. The EMIR E090 receiver was used with the Fourier transform spectrometer backend (FTS) with a spectral resolution of 50 kHz. For the HN$^{13}$C 1-0 transition we used the VESPA backend with a spectral resolution of 6 kHz. The mapping was carried out in good weather conditions ($\tau_{225GHz}$ $\sim$ 0.3) and a typical system temperature of T$_{sys}$ $\sim$ 90-150 K. The data processing was done using the GILDAS software \citep{pet05}. The emission maps have a beam size of 30.1$^{\prime\prime}$, and were gridded to a pixel size of 6$''$ with the CLASS software in the GILDAS package, which corresponds to $\sim$1/5 of the beam size, chosen to ensure Nyquist sampling. The integrated intensity maps of H$^{13}$CN and HC$^{15}$N shown in Figure~\ref{fig:ii_maps} have been computed in the 6.8-7.8 km s$^{-1}$ velocity range, where the v$_{LSR}$ of the source is 7.2 km s$^{-1}$. The integrated emission maps of H$^{13}$CN and HC$^{15}$N are shown in Figure~\ref{fig:ii_maps}. Given the weakness of the lines, we could not produce maps for $^{13}$CN and C$^{15}$N. We were nevertheless able to extract their spectra towards different regions across L1544, and they are shown in Section~\ref{analysis}.

\begin{figure}
\begin{center}
\includegraphics[width=7cm]{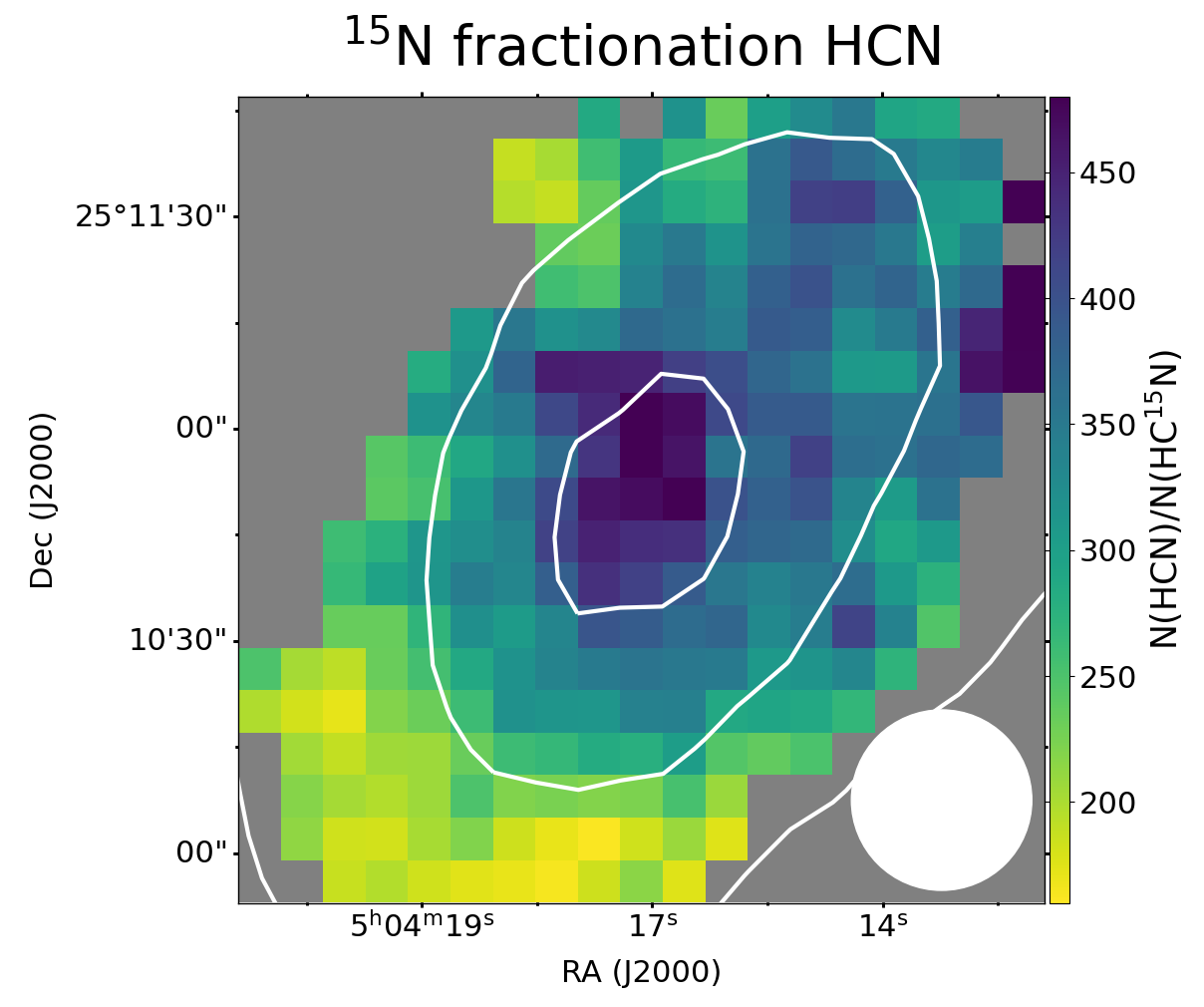}
\end{center}
\caption{$^{14}$N/$^{15}$N ratio map of HCN towards L1544. The map was computed only in the pixels where the integrated emission of both H$^{13}$CN and HC$^{15}$N was detected with a signal-to-noise ratio larger than 5. The IRAM 30m beam of 30$''$ ($\sim$5000 au) are shown in the bottom right of the map. The solid white contours are the 30\%, 60\% and 90\% of the peak intensity of the N(H$_2$) map of L1544 computed from $Herschel$/SPIRE data \citep{spezzano16}. The column density of HCN was computed from the column density of H$^{13}$CN assuming the $^{12}$C/$^{13}$C ratio of 68 \citep{milam05}. The corresponding error is shown in Figure~\ref{fig:15N_maps_error}. }
\label{fig:15N_maps}
\end{figure}

\begin{figure*}
\begin{center}
\includegraphics[width=19cm]{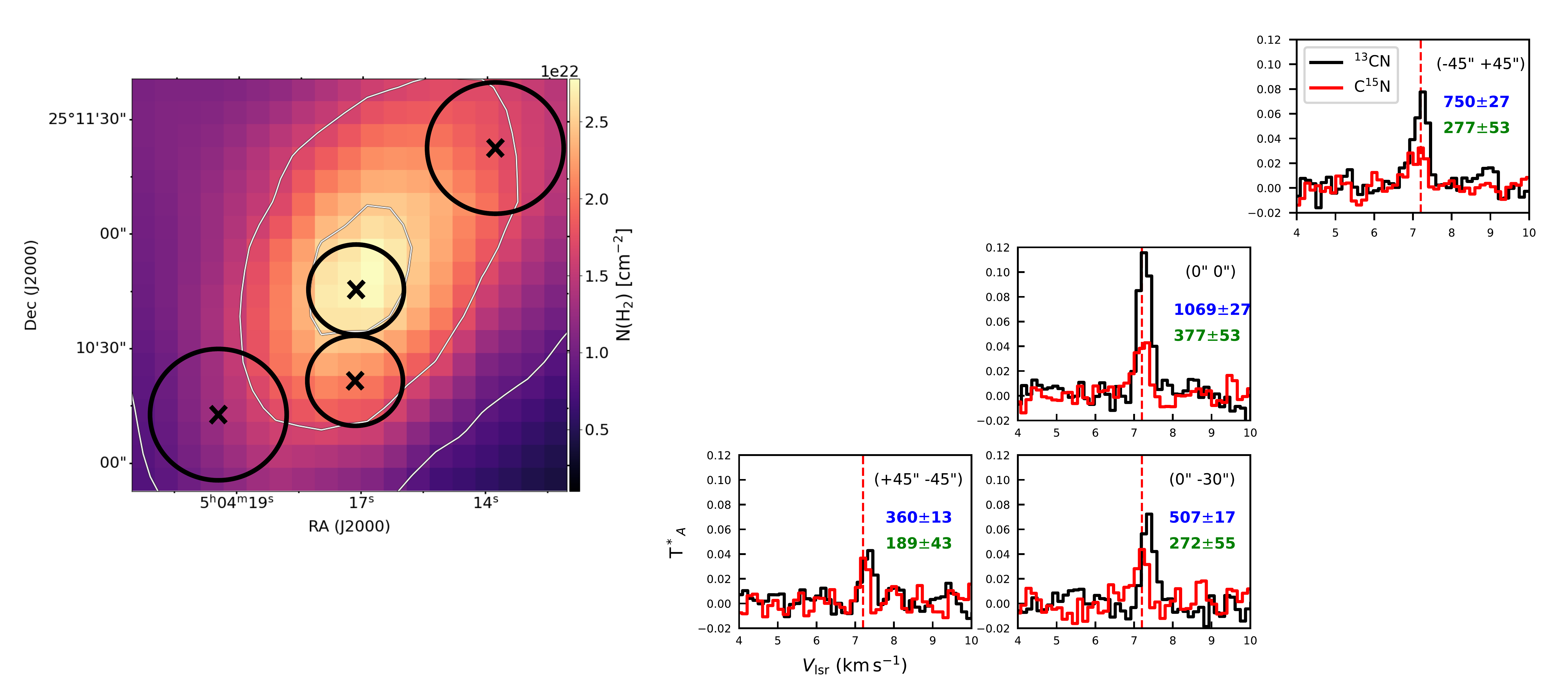}
\end{center}
\caption{Left panel: H$_2$ column density map of L1544 computed from $Herschel$/SPIRE observations \citep{spezzano16}. The black crosses show the centre of the areas where the spectra shown in the right panel have been extracted. The black circles show the regions where the spectra shown in the right panel have been averaged. Right panel: Spectra of $^{13}$CN ($N, J, F_1, F$ =1 3/2 2 3 - 0 1/2 1 2) in black and C$^{15}$N ($N, J, F$= 1 3/2 2 - 0 1/2 1) in red, extracted towards the offsets marked in the left panel. The values in boldface in blue report the $^{14}$N/$^{15}$N ratio in CN for each set of spectra derived using the $^{12}$CN column density directly from the spectra shown in Figure~\ref{fig:13C_spectra}. The values in boldface in green report the $^{14}$N/$^{15}$N ratio computed from the column density of $^{13}$CN using the $^{12}$C/$^{13}$C ratio of 68 \citep{milam05}.}
\label{fig:CN_spectra}
\end{figure*}

\section{Analysis}
\label{analysis}
The spectrum of the 1-0 transition of H$^{13}$CN observed toward the dust peak of L1544 was extracted from the centre of the map shown in Figure~\ref{fig:ii_maps} and fit with the Hyperfine Structure (HFS) tool in \textsc{class} using frequencies measured in the laboratory \citep{fuchs04} with the assumption of the same excitation temperature of the hyperfine components. We derived T$_{ex}$= 3.5 K and an optical depth of 0.4, 2 and 1.2 for the $J$ = 1-0 $F$ = 1-1, 2-1, and 0-1 components, respectively. We repeated the same exercise towards a sample of 4 other positions in the mapped area, to test whether we could assume a constant 3.5 K T$_{ex}$ across L1544, which was the case. The integrated intensity maps shown in Figure~\ref{fig:ii_maps} have been used to compute the column density maps of H$^{13}$CN and HC$^{15}$N assuming T$_{ex}$ = 3.5 K, and subsequently the $^{14}$N/$^{15}$N ratio map shown in Figure~\ref{fig:15N_maps}. In order to retrieve the column density of the main isotopologue, we have multiplied the map of H$^{13}$CN with the isotopic ratio for the local interstellar medium, $^{12}$C/$^{13}$C = 68 \citep{milam05}.
The column density maps have been computed using the formula reported in \cite{Mangum15} for the optically thick emission, assuming that the source fills the beam, and with a constant T$_{ex}$ = 3.5 K:
 
\begin{equation} 
 N_{tot} = \frac{8\pi\nu^3Q_{rot}(T_{ex})W}{c^3A_{ul}g_u}\frac{e^{\frac{E_u}{kT}}}{J(T_{ex}) -  J(T_{bg})}\frac{\tau}{1-e^{-\tau}},
\end{equation}
with
\begin{equation} 
 \tau = \ln\left(\frac{J(T_{ex}) -  J(T_{bg})}{J(T_{ex}) - J(T_{bg}) - T_{mb}}\right).\\
\end{equation}

\noindent where $\frac{\tau}{1-e^{-\tau}}$ is the optical depth correction factor, $\tau$ is the line opacity, $J(T) = {\frac{h\nu}{k}}(e^{\frac{h\nu}{kT}}-1)^{-1}$ is the equivalent Reyleigh-Jeans temperature, $k$ is the Boltzmann constant, $\nu$
is the frequency of the line, $h$ is the Planck constant, $c$ is the speed
of light, $A_{ul}$  is the Einstein
coefficient of the transition, $g_u$ is the degeneracy of the upper state, $E_u$ is the
energy of the upper state, and $Q_{rot}$ is the partition function of the molecule at the given temperature $T_{ex}$. $T_{bg}$, $T_{mb}$ are the
background (2.7 K) and the main beam temperatures respectively. The maps shown in Figure~\ref{fig:15N_maps} have been computed only in the pixels where the integrated emission of both molecules was detected with a signal-to-noise ratio larger than 5.

The emission maps of $^{13}$CN and C$^{15}$N do not have a signal-to-noise ratio that allows for the computation of the $^{14}$N/$^{15}$N ratio map. To check whether the same trend observed in HCN is also found in CN, we averaged the C$^{15}$N 1-0 transition and one of the hyperfine components of the $^{13}$CN 1-0 transition towards several positions across the core. In the left panel of Figure~\ref{fig:CN_spectra} the regions where the spectra of C$^{15}$N and $^{13}$CN have been averaged are marked as black circles on the H$_2$ column density map of L1544. The size of the areas has been optimised to have a signal-to-noise ratio of at least 3. The black crosses mark the center of each region, and their offsets with respect to the center of the map (i.e. the dust peak of L1544) are reported in the respective spectra shown in the right panel. The right panel of Figure~\ref{fig:CN_spectra} shows the averaged spectra from each region.

The hyperfine structure of the 1-0 transition of HN$^{13}$C could not be resolved with the 6 kHz resolution of the VESPA spectra. 
The line shape that we observe deviates from previous observations towards other starless cores, e.g. \cite{vandertak09, padovani11}, suggesting that some of the hyperfine transitions might be self-absorbed. We used the four effective hyperfine transitions described in \cite{vandertak09} and derived the optical depth of each component using the HFS tool in \textsc{class}. The resulting $\tau$ are 1.7, 3.1, 1.9 and 0.1, for the four lines ordered by increasing velocity. Given that the weakest hfs component, at 87090.67 MHz, is optically thin towards the dust peak, we performed a gaussian fit with 4 components for all pixels in our map with \textsc{pyspeckit} \citep{pyspeckit}, with the aim of isolating the thin component and using it to compute the column density map. Unfortunately, the resulting integrated intensity map is observed at a 3$\sigma$ only in the central 30$''$ of the map.  
We extracted the spectra of HN$^{13}$C and H$^{15}$NC towards the same positions used for the CN isotopologues in Figure~\ref{fig:CN_spectra}, see Figure~\ref{fig:HNC_spectra}. To make a direct comparison also with HCN, we have extracted the spectra of the $^{13}$C and $^{15}$N isotopologues of HCN towards the same regions, see Figure~\ref{fig:HCN_spectra}.

\section{Results and discussion}
The $^{14}$N/$^{15}$N ratio map of HCN in the left panel of Figure~\ref{fig:15N_maps} shows a clear decrease towards the South-East of L1544. The HC$^{14}$N/HC$^{15}$N in the North-West of the core is 367$\pm$54, while in the South-East it is 187$\pm$34. Towards the dust peak the HC$^{14}$N/HC$^{15}$N 437$\pm$63, similar to the $^{14}$N/$^{15}$N ratio reported for HC$_3$N in \citealt{hily-blant18}.
The pre-stellar core L1544 is located at the end of a filament in the eastern edge of the Taurus Molecular Cloud. Because of its location and its structure, the southern part of L1544 is more efficiently illuminated by the interstellar radiation field (ISRF) than the northern part, and this has already been shown to have an impact on the chemical differentiation within the core \citep{spezzano16}. 
The $^{14}$N/$^{15}$N ratio map of HCN towards L1544 in this work suggests that the uneven illumination from the ISRF on L1544 has an impact on the $^{15}$N fractionation. As the largest $^{15}$N fractionation is observed toward the Southern part of L1544, the most illuminated by the ISRF, we conclude that the dominant fractionation process is the isotope-selective photodissociation of N$_2$ \citep{heays14}. \cite{guzman17} tentatively observed the effect of isotope-selective photodissociation towards the protoplanetary disk V4046 Sgr. This result was later confirmed by \cite{hily-blant19} towards the disk orbiting the T Tauri star TW Hya. 
The $^{14}$N/$^{15}$N ratio in HCN increases towards the dust peak of L1544 because we are looking through the densest regions of L1544, where the high density reduces the efficiency of the photodissociation, and consequently the enrichment of $^{15}$N in molecular species.

While we cannot compute the $^{14}$N/$^{15}$N ratio maps for CN, the spectra extracted towards four positions across the core in Figure~\ref{fig:CN_spectra} strongly suggest that also for CN the nitrogen fractionation is affected by the ISRF.
In boldface in each spectrum is reported the $^{14}$N/$^{15}$N ratio for CN derived in the corresponding region. In green we report the  C$^{14}$N/C$^{15}$N derived using the $^{12}$C/$^{13}$C ratio of 68, as we did for HCN, and the resulting $^{14}$N/$^{15}$N ratio is similar for both molecules. In blue instead we report the  C$^{14}$N/C$^{15}$N derived using directly $^{12}$CN towards those positions, see Appendix~\ref{13C} for more information. The $^{12}$C/$^{13}$C ratio observed for CN ranges from 120 to 190 and the correspondent C$^{14}$N/C$^{15}$N ratios are as high as 1100, similar to what observed for N$_2$H$^+$ in L1544 and other starless cores \citep{bizzocchi13, redaelli18}. The large $^{12}$C/$^{13}$C and $^{14}$N/$^{15}$N ratios that we present here for CN warrant a dedicated in-depth study of the fractionation of carbon and nitrogen in L1544. It is however important to note that $^{12}$CN/$^{13}$CN ratios larger than 68 have already been observed towards L1544 in \cite{hily-blant10}, and are predicted by the chemical models in \cite{roueff15} and \cite{colzi20}, although for either early time steps or low volume densities. For the purpose of this Letter we highlight that our conclusions on the effect of the illumination on the C$^{14}$N/C$^{15}$N are not affected by the choice of $^{12}$C/$^{13}$C ratio. We do not use the $^{12}$C/$^{13}$C ratio derived from CN for HCN, because the two molecules do not necessarily share all fractionation pathways, as shown in Figures 6 and 7 of \cite{colzi20}.

Figure~\ref{fig:HNC_spectra} shows the spectra of the 1-0 transitions of HN$^{13}$C and H$^{15}$NC extracted in the same position used for CN in Figure~\ref{fig:CN_spectra} and for HCN in Figure~\ref{fig:HCN_spectra}. We computed the HN$^{13}$C column density using the thin hyperfine component of the 1-0 transition following the description of the four effective components described in \cite{vandertak09}, the $^{12}$C/$^{13}$C ratio of 68, and T$_{ex}$ = 4.5 K derived from the \textsc{class} HFS fit towards the dust peak. In boldface in each spectrum is reported the $^{14}$N/$^{15}$N ratio for HNC derived in the corresponding region using the $^{12}$C/$^{13}$C ratio of 68. The resulting $^{14}$N/$^{15}$N ratios are significantly different with respect to HCN and CN, and the trend across the core shows an increase of the $^{14}$N/$^{15}$N ratio towards the South-East, contrary to what we observe for HCN and CN. 
While spatial variations of the fractionation of HCN, CN and HNC across a core have not been explored by chemical models yet, Figure 8 in \cite{roueff15} show that their $^{14}$N/$^{15}$N abundance ratio profiles can differ, especially between 10$^5$ and 10$^6$ yr. Nevertheless, the $^{14}$N/$^{15}$N ratios in HCN, CN and HNC shown in Figure 8 of \cite{roueff15} only range between 390 and 450, while we observe ratios that vary from 150 to 450.  Non-LTE modelling is necessary to confirm our results, in particular for the HN$^{13}$C line where an isolated hyperfine component is not present.

\section{Conclusions}
Our $^{14}$N/$^{15}$N ratio map of HCN towards L1544 shows for the first time that the fractionation of nitrogen presents significant variations across a pre-stellar core.
The $^{14}$N/$^{15}$N ratio in HCN decreases towards the South-East of the core, the region of L1544 that corresponds to a steeper drop in H$_2$ column density and is consequently more efficiently illuminated by the ISRF. This was already shown in previous observations of carbon-chain molecules which in fact peak toward the region L1544 more exposed to the ISRF, where a significant fraction of carbon is maintained in atomic form \citep{spezzano16}.
The same trend is observed also for CN and the opposite trend is observed for HNC.
Our results indicate that isotope-selective photodissociation plays an important role in the fractionation of nitrogen in L1544. $^{14}$N$^{15}$N photodissociates more efficiently than $^{14}$N$_2$ because it is not abundant enough to self-shield. The photodissociation of $^{14}$N$^{15}$N is expected to be more efficient towards the more illuminated southern part of the core, where more atomic $^{15}$N will be available to form cyanides like HCN and CN. HNC shows opposite behaviour with respect to HCN and CN. Further studies are necessary to understand the underlying cause.

The effect of isotope-selective photodissociation in nitrogen fractionation has already been observed towards a protoplanetary disk where the irradiation from UV photons in the inner part of the disk translates into a lower $^{14}$N/$^{15}$N ratio in HCN \citep{hily-blant19}. With our work we show that the uneven illumination from the ISRF onto a pre-stellar core has an effect on the $^{14}$N/$^{15}$N ratio through the isotope-selective photodissociation. With the $^{14}$N/$^{15}$N ratio in atomic nitrogen decreasing towards the southern part of L1544, the $^{14}$N$_2$/$^{14}$N$^{15}$N ratio will have the opposite behaviour and increase towards the South of L1544, because the $^{14}$N$_2$ is expected to be less affected by photodissociation. As a consequence, we expect the $^{14}$N/$^{15}$N ratio in molecules that are formed from molecular nitrogen, like N$_2$H$^+$, to show the opposite behaviour with respect to HCN and CN. This trend has already been observed towards the high-mass star forming region IRAS 05358+3543 in \cite{colzi19}. Future maps of $^{14}$N$_2$H$^+$/$^{14}$N$^{15}$NH$^+$ or $^{14}$N$_2$H$^+$/$^{15}$N$^{14}$NH$^+$ towards L1544 are needed to confirm this point for low-mass star-forming regions. 

We also present here the direct measurement of the $^{12}$CN/$^{13}$CN ratio across L1544, showing that it ranges between 130 and 190 in the inner 2$\times$2 arcmin$^2$ of the core. Furthermore, we show that column density ratios between the $^{13}$C and $^{15}$N isotopologues is  $\sim$6 for HCN and CN and $\sim$2 for HNC towards the dust peak. A dedicated work on the chemical modelling of carbon and nitrogen fractionation towards L1544 in currently ongoing. Nevertheless, our results on CN indicate that the effect of the illumination on the nitrogen fractionation of HCN and CN across the core is not affected by the fractionation of carbon.  

Pre-stellar cores provide the budget of material that will finally be inherited by forming planets. In order to assess what is the $^{14}$N/$^{15}$N budget that can be inherited from pre-stellar cores, it is important to consider the illumination-induced variations across the core, which could account for different $^{15}$N reservoirs of future stellar systems, together with the $^{15}$N fractionation variation within a protoplanetary disk \citep{hily-blant17,hily-blant19}.

\begin{acknowledgements}
The authors wish to thank Laura Colzi for useful discussion, and the anonymous referee for insightful suggestions.
We gratefully acknowledge the support of the Max Planck Society.
\end{acknowledgements}

%
%

{}

\begin{appendix}

\section{Observed spectra}
\begin{figure*}

\begin{center}
\includegraphics[width=17cm]{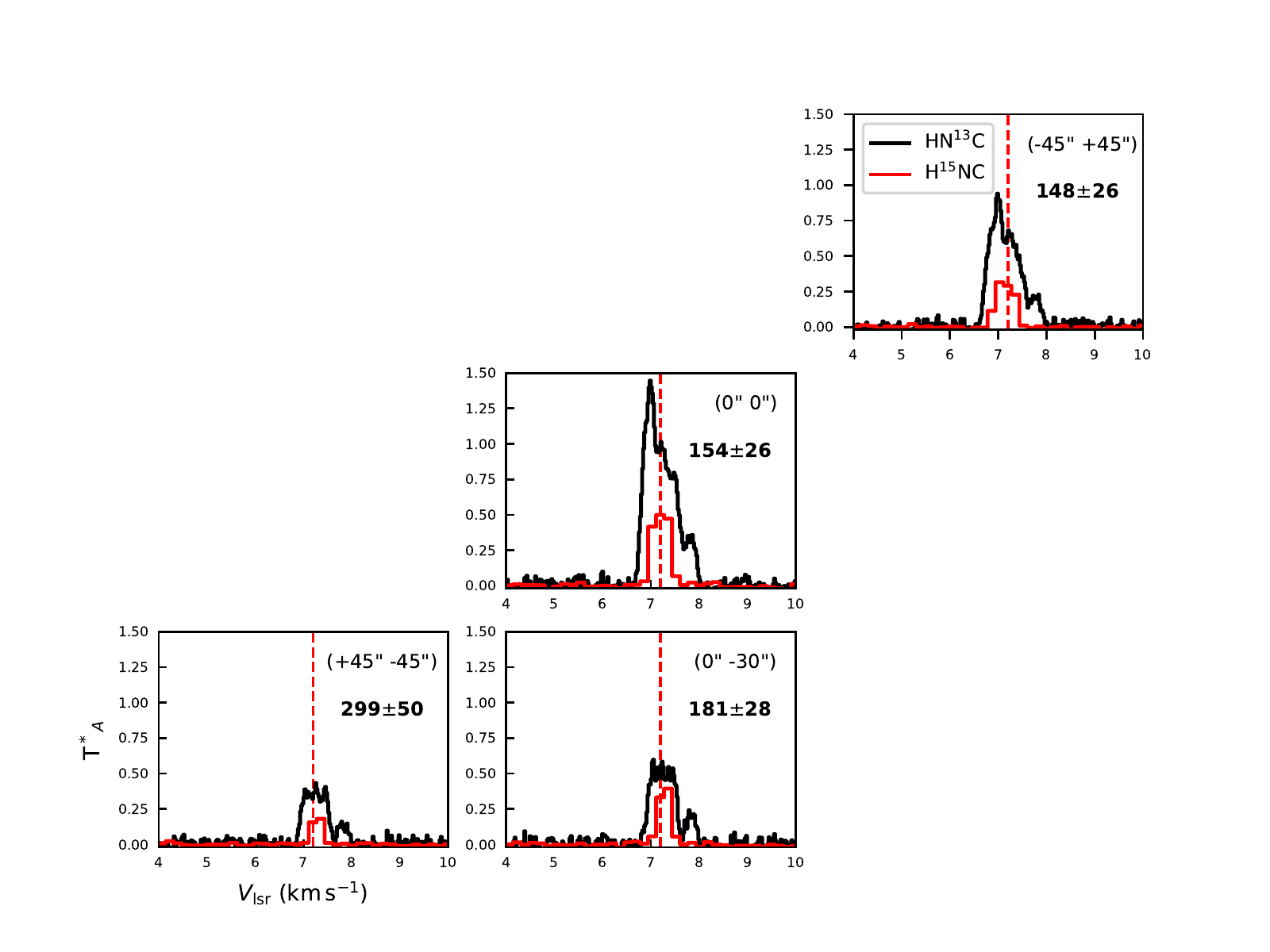}
\end{center}
\caption{Spectra of HN$^{13}$C ($N$ = 1-0) in black and H$^{15}$NC ($N$= 1-0) in red extracted in the regions shown as black circles in the left panel in Figure~\ref{fig:CN_spectra}. The values in boldface report the $^{14}$N/$^{15}$N ratio in HNC for each set of spectra derived using the $^{12}$C/$^{13}$C ratio of 68 \citep{milam05}.} 
\label{fig:HNC_spectra}
\end{figure*}

\begin{figure*}

\begin{center}
\includegraphics[width=17cm]{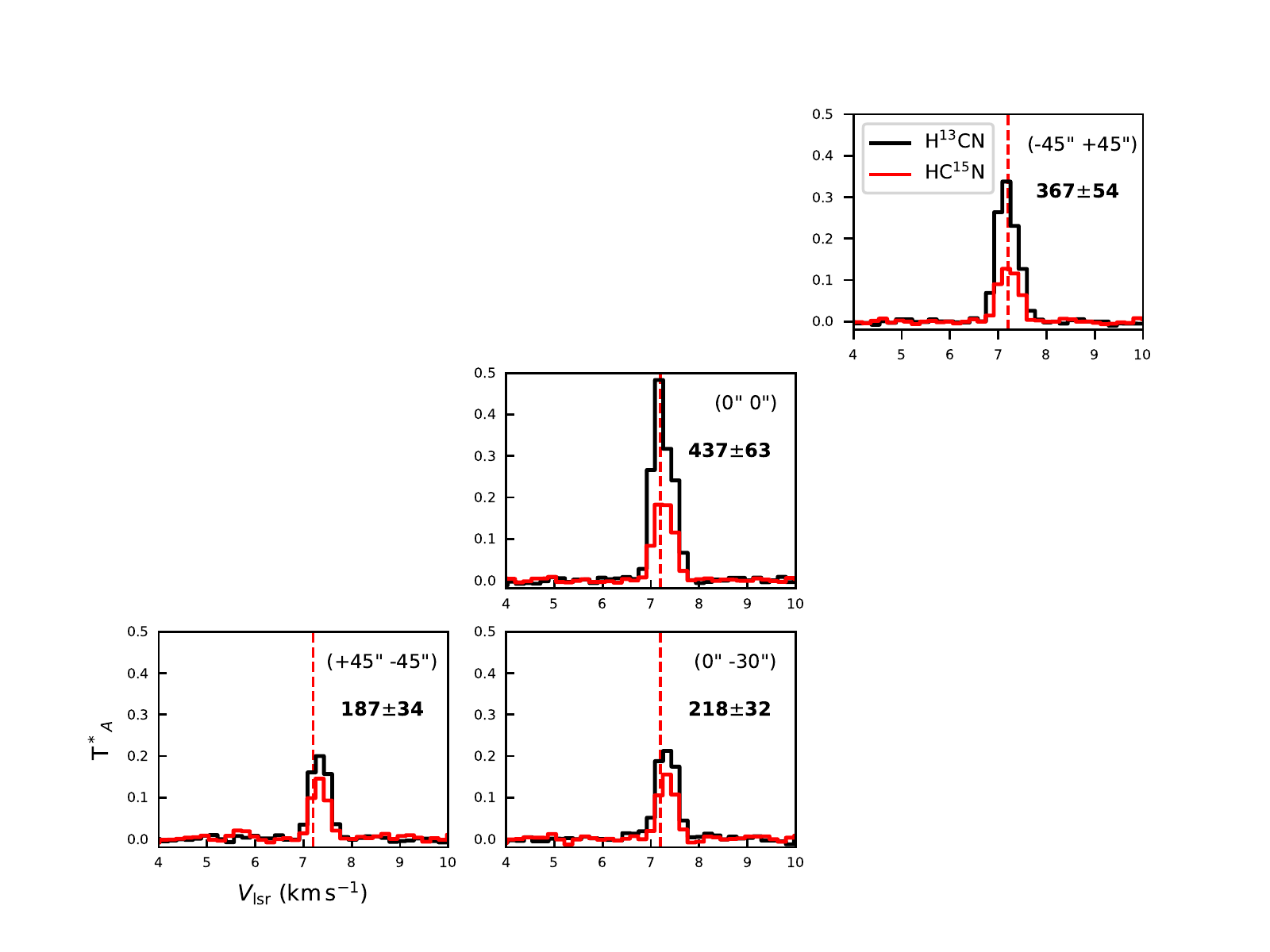}
\end{center}
\caption{Spectra of H$^{13}$CN ($J, F$ = 1 2 - 0 1) in black and HC$^{15}$N ($N$= 1-0) in red extracted in the regions shown as black circles in the left panel in Figure~\ref{fig:CN_spectra}. The values in boldface report the $^{14}$N/$^{15}$N ratio in HCN for each set of spectra derived using the $^{12}$C/$^{13}$C ratio of 68 \citep{milam05}. Note the sharp decrease of the H$^{13}$CN line toward the South, when compared to the almost constant HC$^{15}$N.}
\label{fig:HCN_spectra}
\end{figure*}

Figures~\ref{fig:HNC_spectra} and \ref{fig:HCN_spectra} show the spectra of the $^{13}$C and $^{15}$N isotopologues of HCN and HNC towards the positions shown in the left panel of Figure~\ref{fig:CN_spectra} (where the spectra of $^{13}$CN and C$^{15}$N were extracted). A sharp decrease can be seen towards the South in the lines of the $^{13}$C isotopologues, while the intensities of the lines of the $^{15}$N isotopologues are almost constant across the core.

The spectra shown in Figures~\ref{fig:HNC_spectra} and \ref{fig:HCN_spectra}, as well as in Figure~\ref{fig:CN_spectra} show a small shift in velocity across the core. Such shift is predicted by the model from \cite{ciolek00} shown in Fig. 6 in \cite{caselli02}, and has already been observed in L1544, see for example Figure B.1 in \cite{spezzano16}.

\begin{figure}
\begin{center}
\includegraphics[width=7cm]{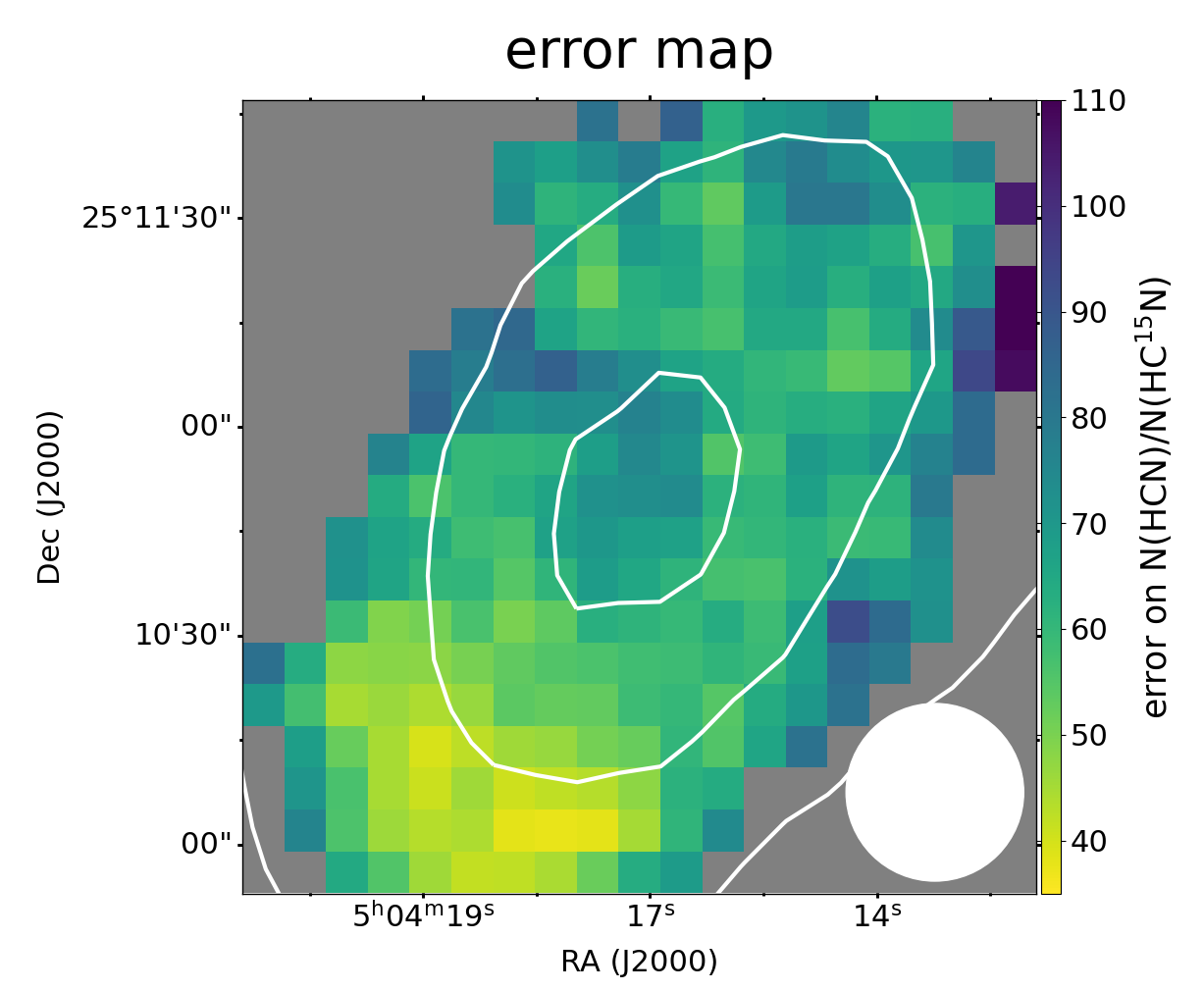}
\end{center}
\caption{Error map of $^{14}$N/$^{15}$N ratio of HCN towards L1544. The map was computed only in the pixels where the integrated emission of both H$^{13}$CN and HC$^{15}$N was detected with a signal-to-noise ratio larger than 5. The IRAM 30m beam of 30$''$ ($\sim$5000 au) are shown in the bottom right of the map. The solid white contours are the 30\%, 60\% and 90\% of the peak intensity of the N(H$_2$) map of L1544 computed from $Herschel$/SPIRE data \citep{spezzano16}.}
\label{fig:15N_maps_error}
\end{figure}

Figure~\ref{fig:15N_maps_error} shows the error map on the HC$^{14}$N/HC$^{15}$N column density ratio map, and is calculated by propagating the errors on the column density maps of H$^{13}$CN and HC$^{15}$N, which include the rms of the spectra as well as a 10\% calibration error.

\section{Effect of the $^{12}$C/$^{13}$C fractionation}
\label{13C}
\begin{figure*}

\begin{center}
\includegraphics[width=17cm]{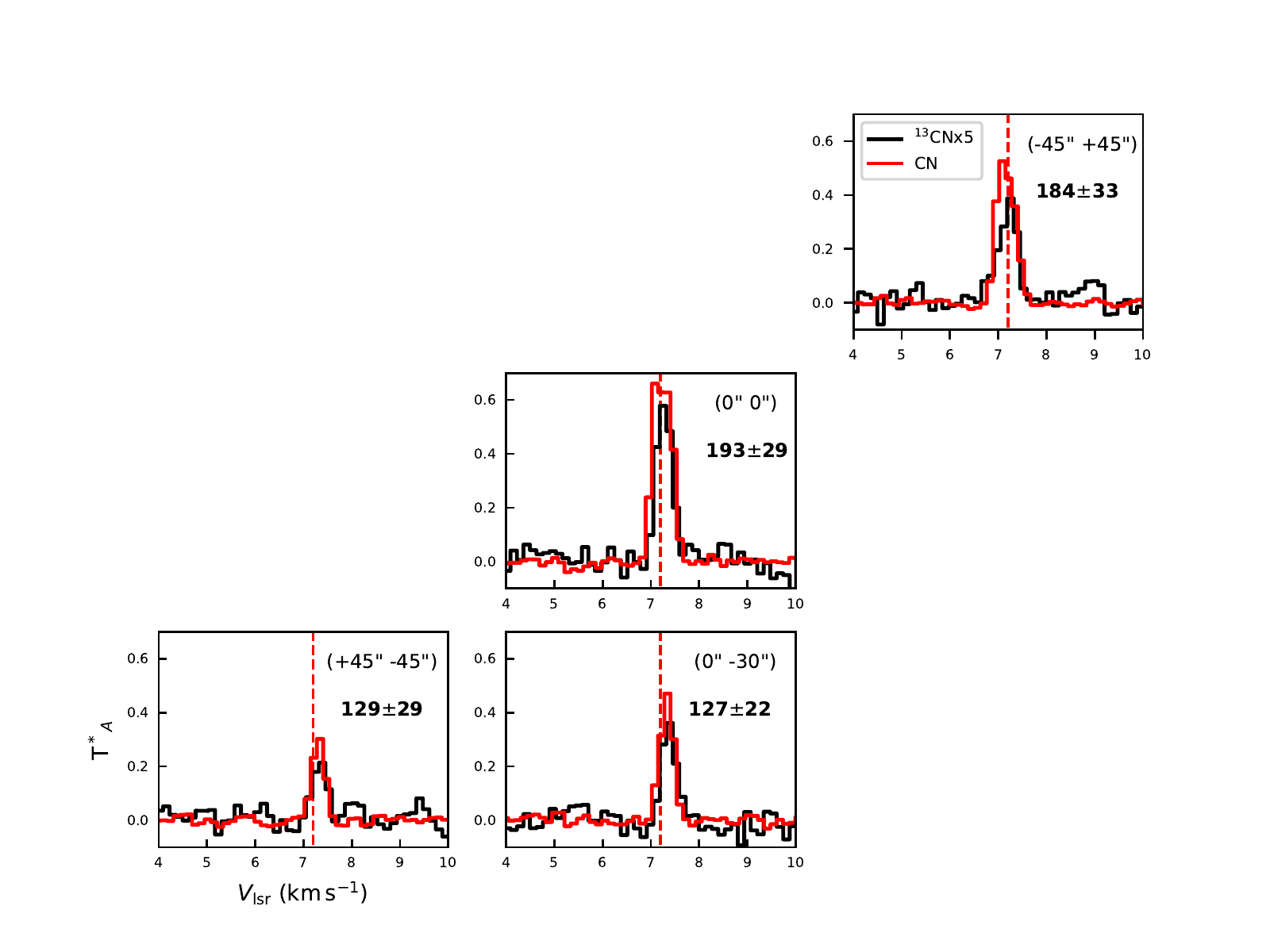}
\end{center}
\caption{Spectra of $^{13}$CN ($N$ =1-0 $J$ = 3/2-1/2 $F_1$ = 2-1 $F$ = 3-2) in black and CN ( $N$ =1-0 $J$ = 1/2-1/2 $F$ = 1-1) in red extracted in the regions shown as black circles in the left panel in Figure~\ref{fig:CN_spectra}. The $^{12}$CN/$^{13}$CN column density ratio derived towards the four offsets is written in boldface in each spectrum. }
\label{fig:13C_spectra}
\end{figure*}

We extracted the spectra of the $N$ =1-0 $J$ = 3/2-1/2 $F_1$ = 2-1 $F$ = 3-2 transition of $^{13}$CN and the $N$ =1-0 $J$ = 1/2-1/2 $F$ = 1-1 of CN towards the four position marked in the left panel in Figure~\ref{fig:CN_spectra} and derived the $^{12}$CN/$^{13}$CN, see Figure~\ref{fig:13C_spectra}. We used the map of the weakest hyperfine component of the 1-0 transition of the main isotopologue, which is only slightly optically thick, with $\tau$ ranging from 1.1 to 1.4 across the map, and does not show signs of self-absorption in our spectra with 50 kHz resolution. 
In contrast to the spectra shown in Figures~\ref{fig:HCN_spectra} and \ref{fig:HNC_spectra}, where the lines of the $^{13}$C and $^{15}$N isotopologues do not decrease at the same pace towards the south of L1544, the intensity variations of the lines of $^{12}$CN and $^{13}$CN do not show a substantial difference. We computed the column density for $^{12}$CN and $^{13}$CN from the spectra shown in Figure~\ref{fig:13C_spectra}, assuming optically thick emission, and T$_{ex}$=4.2K. The excitation temperature and optical depth were derived by modelling the observed line of $^{12}$CN with RADEX assuming a kinetic temperature of 8 K and a volume density of 1$\times$10$^5$ cm$^{-3}$, and is consistent with the excitation temperature previously derived for $^{13}$CN and C$^{15}$N in \cite{hily-blant13a}. Please note that the choice of volume density has an impact on the resulting excitation temperature, and consequently on the optical depth of the $^{12}$CN line and the resulting column density. For example, if we assume a volume density of 5$\times$10$^5$ cm$^{-3}$, the corresponding excitation temperature is 5 K, which reduces the optical depth of the $^{12}$CN line by almost a factor of two, and in turn decreases the $^{12}$CN/$^{13}$CN ratios in the central and northern offset, where the $^{12}$CN line is brighter.
The $^{12}$CN/$^{13}$CN ratio derived in this work (193$\pm$10 towards the dust peak) is larger than the values derived in B1b, where $^{12}$CN/$^{13}$CN = 50$^{19}_{11}$ \citep{daniel13}. However, the chemical models presented in \cite{colzi20} can reproduce values of $^{12}$CN/$^{13}$CN larger than 68 (the isotopic ratio for the local interstellar medium, \citealt{milam05}), as seen in Figure 7 of \cite{colzi20}. The variation of the $^{12}$CN/$^{13}$CN ratio with density and time in the models are mainly connected to the competition between the enrichment of carbon monoxide in $^{13}$C and the availability of $^{13}$C$^+$. 
Our results on CN indicate that the effect of the illumination on the nitrogen fractionation across the core is not affected by the fractionation of carbon. Consequently, we can assume that the $^{14}$N/$^{15}$N ratio map of HCN in the left panel of Figure~\ref{fig:15N_maps} might need to be corrected by an offset (i.e. a different $^{12}$C/$^{13}$C ratio), but it will not change its trend substantially because of the carbon fractionation.
Chemical modeling for the combined fractionation of carbon and nitrogen are currently undergoing and go beyond the scope of this Letter. It is however important to note that the ratio between the $^{13}$C and $^{15}$N isotopologues for CN and HCN is the same towards the dust peak of L1544, $\sim$6, and drops for both molecules to $\sim$3 towards the South-West of the core, strongly suggesting that both molecules undergo the same chemical paths for the fractionation of Nitrogen. While the normal isotopologues trace the core and the cloud where the core is embedded, it is safe to assume that the $^{13}$C and $^{15}$N isotopologues trace only the core. Therefore, it might be more appropriate to compare the $^{13}$C and $^{15}$N isotopologue ratios instead of the ratios involving the main isotopologue, even when it is possible to derive a direct measurement.

\end{appendix}

\end{document}